# Using Cloud-Aware Provenance to Reproduce Scientific Workflow Execution on Cloud

Khawar Hasham[1], Kamran Munir[1] and Richard McClatchey[1]

[1]Centre for Complex Computing Systems (CCCS), Faculty of Environment and Technology (FET),
University of the West of England (UWE), Bristol, United Kingdom
{khawar.ahmad, kamran.munir}@cern.ch,

Keywords: Cloud computing, Scientific Workflows, Provenance, Reproducibility, Repeatability

Abstract: Provenance has been thought of a mechanism to verify a workflow and to provide workflow reproducibility. This provenance of scientific workflows has been effectively carried out in Grid based scientific workflow systems. However, recent adoption of Cloud-based scientific workflows present an opportunity to investigate the suitability of existing approaches or propose new approaches to collect provenance information from the Cloud and to utilize it for workflow repeatability in the Cloud infrastructure. This paper presents a novel approach that can assist in mitigating this challenge. This approach can collect Cloud infrastructure information from an outside Cloud client along with workflow provenance and can establish a mapping between them. This mapping is later used to re-provision resources on the Cloud for workflow execution. The reproducibility of the workflow execution is performed by: (a) capturing the Cloud infrastructure information (virtual machine configuration) along with the workflow provenance, (b) re-provisioning the similar resources on the Cloud and re-executing the workflow on them and (c) by comparing the outputs of workflows. The evaluation of the prototype suggests that the proposed approach is feasible and can be investigated further. Moreover, there is no reference reproducibility model exists in literature that can provide guidelines to achieve this goal in Cloud. This paper also attempts to present a model that is used in the proposed design to achieve workflow reproducibility in the Cloud environment.

## 1 INTRODUCTION

The scientific community is processing and analysing huge amounts of data being generated in modern scientific experiments that include projects such as DNA analysis (Foster 2008), the Large Hadron Collider (LHC) (http://lhc.cern.ch), and projects such as neuGRID (Mehmood 2009) and its follow-on neuGRIDforUsers (Munir 2013, 2014). In particular the neuGRID community is utilising scientific workflows to orchestrate the complex analysis of neuro-images to diagnose Alzheimer disease. A large pool of compute and data resources are required to process this data, which has been available through the Grid (Foster 1998) and is now also being offered by the Cloud-based infrastructures.

Cloud computing (Mell 2009) has emerged as a new computing and storage paradigm, which is dynamically scalable and usually works on a pay-as-you-go cost model. It aims to share resources to store data and to host services transparently among users at a massive scale (Mei 2008). Its ability to provide an on-demand computing infrastructure enables distributed processing of scientific workflows (Deelman 2008) with increased complexity and data requirements. Recent work (Juve 2008) is now experimenting with Cloud infrastructures to assess the feasibility of executing workflows on the Cloud.

An important consideration during this data processing is to gather provenance (Simmhan 2005) information that can provide detailed information about both the input and the processed output data, and the processes involved in a workflow execution. This information can be used to debug the execution of a workflow, to aid in error tracking and reproducibility. This vital information can enable scientists to verify the outputs and iterate on the scientific method, to evaluate the process and results of other experiments and to share their own experiments with other scientists (Azarnoosh 2013).

The execution of scientific workflows in Cloud brings to the fore the need to collect provenance information that is necessary to ensure the reproducibility of these experiments on Cloud infrastructure

A research study (Zhao 2012) conducted to evaluate the reproducibility of scientific workflows has shown that around 80% of the workflows cannot be reproduced, and 12% of them are due to the lack of information about the execution environment. This information affects a workflow on two levels. It can affect a workflow's overall execution performance and also job failure rate. For instance, a data-intensive job can perform better with 2GB of RAM because it can accommodate more data in RAM, which is a faster medium than hard disk. However, the job's performance will degrade if a resource of 1GB RAM is allocated to this job as less data can be placed in RAM. Moreover, it is also possible that jobs will remain in waiting queues or fail during execution if their required hardware dependencies are not met. This becomes more challenging issue in the context of Cloud in which resources can be created or destroyed at runtime.

The dynamic and geographically distributed nature of Cloud computing makes the capturing and processing of provenance information a major research challenge (Vouk 2008, Zhao 2011). Since the Cloud presents a transparent access to dynamic execution resources, the workflow parameters including execution resource configuration should also be known to a scientist (Shamdasani 2012, Cruz 2011) i.e. what execution environment was used for a job in order to reproduce a workflow execution on the Cloud. Due to these reasons, there is a need to capture information about the Cloud infrastructure along with workflow provenance, to aid in the reproducibility of workflow experiments. There has been a lot of research related to provenance in the Grid (Foster 2002, Stevens 2003) and a few initiatives (Oliveira 2010, Ko 2012) for the Cloud. However, they lack the information that can be utilised for re-provisioning of resources on the Cloud, thus they cannot create the similar execution environment(s) for workflow repeatability. In this paper, the terms "Cloud infrastructure" and "virtualization layer" are used interchangeably.

This paper presents a theoretical description of an approach that can augment workflow provenance with infrastructure level information of the Cloud and use it to establish similar execution environment(s) and repeat a given workflow. Important points discussed in this paper are as follows: section 2 presents some related work in provenance related systems. Section 3 presents a reproducibility model designed after collecting guidelines used and discussed in literature. Section 4 presents an overview of the proposed approach. Section 5 presents an evaluation of the developed prototype. And finally section 6 presents some conclusions and directions for future work.

## 2 RELATED WORK

Significant research (Foster 2002, Scheidegger 2008) has been carried out in workflow provenance for Grid-based workflow management systems. Chimera (Foster 2002) is designed to manage the data-intensive analysis for high-energy physics (GriPhyN) (GriPhyN 2014) and astronomy (SDSS) (SDSS 2014) communities. It captures process information, which includes the runtime parameters, input data and the produced data. It stores this provenance information in its schema, which is based on a relational database. Although the schema allows storing the physical location of a machine, it does not support the hardware configuration and software environment in which a job was executed. Vistrails (Scheidegger 2008) provides support for scientific data exploration and visualization. It not only captures the execution log of a workflow but also the changes a user makes to refine his workflow. However, it does not support the Cloud virtualization layer information. Similar is the case with Pegasus/Wings (Kim et al. 2008) that supports evolution of a workflow. However, this paper is focusing on the workflow execution provenance on the Cloud, rather than the provenance of a workflow itself (e.g. design changes).

There have been a few research studies (Oliveira 2010, Ko 2011) performed to capture provenance in the Cloud. However, they lack the support for workflow reproducibility. Some of the work in Cloud towards provenance is directed to the file system (Zhang 2011, Shyang et al 2012) or hypervisor level (Macko 2013). However this work is not relatable to our approach because this paper focuses on virtualized layer information of the Cloud for workflow execution. Moreover, the collected provenance data provides information about the file access but it does not provide information about the resource configuration. The PRECIP (Azarnoosh 2013) project provides an API to provision and execute workflows. However, it does not provide provenance information of a workflow.

There have been a few recent projects (Chirigati 2013, Yves 2014) and research studies (Perez

2014a) on collecting provenance and using it to reproduce an experiment. A semantic-based approach (Perez 2014b) has been proposed to improve reproducibility of workflows in the Cloud. This approach uses ontologies to extract information about the computational environment from the annotations provided by a user. This information is then used to recreate (install or configure) that environment to reproduce a workflow execution. On the contrary, our approach is not relying on annotations rather it directly interacts with the Cloud middleware at runtime to acquire resource configuration information and then establishes mapping between workflow jobs and Cloud resources. The ReproZip software (Chirigati 2013) uses system call traces to provide provenance information for job reproducibility and portability. It can capture and organize files/libraries used by a job. The collected information along with all the used system files are zipped together for portability and reproducibility purposes. Since this approach is useful at individual job level, this does not work for an entire workflow, which is the focus of this paper. Moreover, this approach does not consider the hardware configuration of the underlined execution machine. Similarly, a Linux-based tool, CARE (Yves 2014), is designed to reproduce a job execution. It builds an archive that contains selected executable/binaries and files accessed by a given job during an observation run.

## 3  WORKFLOW REPRODUCI-BILITY MODEL ON CLOUD

As per our understanding of the literature, there is not a standard reproducibility model proposed so far for scientific workflows, especially in Cloud environment. However, there are some guidelines or policies, which have been highlighted in literature to reproduce experiments. There is one good effort (Sandve 2013) in this regard, but it mainly talks about reproducible papers and it does not consider execution environment of workflows. In this section, we have gathered basic points to present an initial workflow reproducibility model in Cloud that can provide guidelines for future work in this regard. These points are discussed as follows.

- Share Code and Data

The need for data and code sharing in computational science has been widely discussed (Stodden 2010). In computational science conservation, in particular for scientific workflow executions, it is emphasized that the data, code, and the workflow description should be available in order to reproduce an experiment.

- Execution Infrastructure details

A workflow is executed on an underlined infrastructure provided by Grid or Cloud. The execution infrastructure is composed of a set of computational resources (e.g. execution nodes, storage devices, networking). The physical approach, where actual computational hardware are made available for long time period to scientists, often conserves the computational environment including supercomputers, clusters, or Grids (Perez 2014b). As a result, scientists are able to reproduce their experiments in the same hardware environment. However, this luxury is not available in Cloud in which resources are virtual and dynamic. This challenge has been tackled in this paper by collecting this information at runtime from the Cloud infrastructure.

- Software Environment

Apart from knowing the hardware infrastructure, it is also essential to provide information about the software environment. A software environment determines the operating system and the libraries used to execute a job. Without the access to required libraries information, a job execution will fail. For example, a job, relying on MATLAB library, will fail in case the required library is missing. One possible approach (Howe 2012) to conserve software environment is thought to conserve VM that is used to execute a job and then reuse the same VM while re-executing the same job. One may argue that it would be easier to keep and share VM images with the research community through a common repository, however the high storage demand of VM images remains a challenging problem (Zhao 2013). In the prototype presented in this paper, VM is conserved and thought to present all the software dependencies required for a job execution in a workflow.

- Workflow Versioning

Capturing only a provenance trace is not sufficient to allow a computation to be repeated – a situation known as workflow decay (Roure 2011). The reason is that the provenance systems can store information on how the data was generated, however they do not store copies of the key actors in the computation i.e. workflow, services, data. This paper (Sandve et al. 2013) suggests archiving the exact versions of all programs and enabling version control on all scripts used in an experiment. This is not supported in the presented prototype, but it will be incorporated in next iterations.

- Provenance Comparison

The provenance of workflow should be compared to determine workflow reproducibility. The comparison should be made at different levels; workflow structure, execution infrastructure, and workflow input and output. A brief description of this comparison is given below.

a) Workflow structure should be compared to determine that both workflows are similar. Because it is possible that two workflows are having similar number of jobs but with different job execution order.
b) Execution infrastructure (software environment, resource configuration) used for a workflow execution should also be compared.
c) Comparison of input and output should be made to confirm workflow reproducibility. There could be a scenario that a user repeated a workflow but with different inputs, thus producing different outputs. It is also possible that changes in job or software library result into different workflow output. There are a few approaches (Zhang 2011, Missier et al. 2013), which perform workflow provenance comparison to determine differences in reproduced workflows. The proposed approach in this paper incorporates the workflow output comparison to determine the reproducibility of a workflow.

- Pricing model

This point can be important for experiments in which cost is also a main factor. On Cloud, resources are provisioned on-demand and cost is associated with the resources. This information can help in reproducing an experiment with the same cost as was incurred in earlier execution. This point is not incorporated in the proposed design at the moment.

## 4 CLOUD-AWARE PROVENANCE APPROACH

An abstract view of the proposed architecture is presented in this section. This architecture is designed after evaluating the existing literature and keeping in mind the objectives of this research study. The proposed architecture is inspired by the mechanism used in a paper (Groth 2009) for executing workflows on the Cloud. Figure 1 illustrates the proposed architecture that is used to capture the Cloud infrastructure information and to interlink it with the workflow provenance collected from a workflow management system such as Pegasus. This augmented or extended provenance information compromising of workflow provenance and the Cloud infrastructure information is named as Cloud-aware provenance. The components of this architecture are briefly explained below.

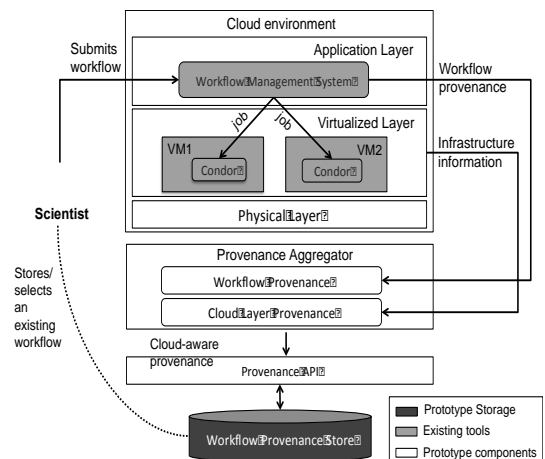

Figure 1: An abstract architecture of the proposed approach.

- **Workflow Provenance:** This component is responsible for receiving provenance captured at the application level by the workflow management system (Pegasus). Since workflow management systems may vary, a plugin-based approach is used for this component. Common interfaces are designed to develop plugins for different workflow management systems. The plugin also translates the workflow provenance according to the representation that is used to interlink the workflow provenance along with the information coming from the Cloud infrastructure.
- **Cloud Layer Provenance**: This component is responsible for capturing information collected from different layers of the Cloud. To achieve re-provisioning of resources on Cloud, this component focuses on the virtualization layer and retrieves information related to the Cloud infrastructure i.e. virtual machine configuration. This component is discussed in detail in section 4.1.
- **Provenance Aggregator:** This is the main component tasked to collect and interlink the provenance coming from different layers as shown in Figure 1. It establishes interlinking connections between the workflow provenance and the Cloud infrastructure information. The provenance information is then represented in a single format that could be stored in the provenance store through the interfaces exposed by the Provenance API.
- **Provenance API**: This acts as a thin layer to expose the provenance storage capabilities to other components. Through its exposed interfaces, outside entities such as the Provenance Aggregator would interact with it to store the workflow provenance information. This approach gives flexibility to implement authentication or authorization in accessing the provenance store.
- **Workflow Provenance Store**: This data store is designed to store workflows and their associated provenance. This also keeps mapping between workflow jobs and the virtual compute resources in the Cloud infrastructure. This also keeps record of the workflow and its related configuration files being used to submit a user analysis on the Cloud. This information is later retrieved to reproduce the execution. However, it does not support workflow evolution in its current design.

## 4.1 Job to Cloud Resource Mapping

The CloudLayerProvenance component is designed in a way that interacts with the Cloud infrastructure as an outside client to obtain the resource configuration information. As mentioned earlier, this information is later used for reprovisioning the resources to provide a similar execution infrastructure to repeat a workflow execution. Once a workflow is executed, Pegasus collects the provenance and stores it in its own internal database. Pegasus also stores the IP address of the virtual machine (VM) where the job is executed. However, it lacks other VM specifications such as RAM, CPUs, hard disk etc. The CloudLayerProvenance component retrieves all the jobs of a workflow and their associated VM IP addresses from the Pegasus database. It then collects a list of virtual machines owned by a respective user from the Cloud middleware. Using the IP address, it establishes a mapping between the job and the resource configuration of the virtual machine used to execute the job. This information i.e. Cloud-aware provenance is then stored in the Provenance Store. The flowchart of this mechanism is presented in Figure 2.

In this flowchart, the variable *wfJobs* – representing a list of jobs of a given workflow – is retrieved from the Pegasus database. The variable *vmList* – represents a list of virtual machines in the Cloud infrastructure – is collected from the Cloud. A mapping between jobs and VMs is established by matching the IP addresses (see in Figure. 2). Resource configuration parameters such as flavour and image are obtained once the mapping is established. *flavour* defines resource configuration such as RAM, Hard disk and CPUs, and *image* defines the operating system image used in that particular resource. By combining these two parameters together, one can provision a resource on the Cloud infrastructure. After retrieving these parameters and jobs, the mapping information is then stored in the Provenance Store (see in Figure. 2). This mapping information provides two important data (a) hardware configuration (b) software configuration using VM name. As discussed in section 3, these two parameters are important in recreating a similar execution environment.

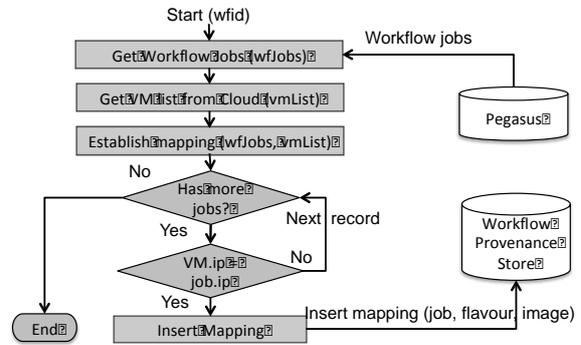

Figure 2: flowchart of job to Cloud resource mapping performed by ProvenanceAggregator component.

## 4.2 Workflow Reproducibility using Cloud-Aware Provenance

In section 4.1, the job to Cloud resource mapping using provenance information has been discussed. This mapping is stored in the database for workflow repeatability purposes. In order to reproduce a workflow execution, researcher first needs to provide the *wfID* (workflow ID), which is assigned to every workflow in Pegasus, to the proposed framework to re-execute the given workflow using the Cloud-aware provenance. It retrieves the given workflow from the Provenance Store database (step 2(a) in Figure 3) along with the Cloud resource mapping stored against this workflow (step 2(b) in Figure 3). Using this mapping information, it retrieves the resource flavour and image configurations, and provisions the resources (step 3 in Figure 3) on Cloud. Once resources are provisioned, it submits the workflow (step 4).

At this stage, a new workflow ID is assigned to this newly submitted workflow. This new wfID is passed over to the ProvenanceAggregator component to monitor (step 5) the execution of the workflow and start collecting its Cloud-aware provenance information (see step 6 in Figure 3) This is important to recollect the provenance of the repeated workflow, as this will enable us to verify the provisioned resources by comparing their resource configurations with the old resource configuration.

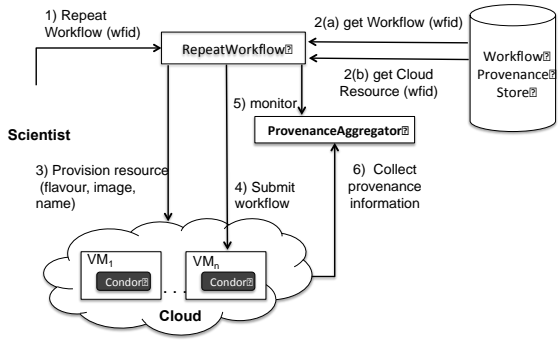

Figure 3: The sequence of activities to illustrate workflow repeatability in the proposed system.

## 4.3 Workflow Output Comparison

Another aspect of workflow repeatability is to verify that it has produced the same output that was produced in its earlier execution (as discussed in section 3). In order to evaluate workflow repeatability, an algorithm has been proposed that compares the outputs produced by two given workflows. It uses the MD5 hashing algorithm (Stalling 2010) on the outputs and compares the hash value to verify the produced outputs. The two main reasons of using a hash function to verify the produced outputs are; a) simple to implement and b) the hash value changes with a single bit change in the file. If the hash values of two given files are same, this means that the given files contain same content.

The proposed algorithm (as shown in Figure 4) operates over the two given workflows identified by *srcWfID* and *destWfID*, and compares their outputs. It first retrieves the list of jobs and their produced output files from the Provenance Store for each given workflow. It then iterates over the files and compares the source file, belonging to *srcWfID*, with the destination file, belonging to *destWfID*. Since the files are stored on the Cloud, the algorithm retrieves the files from the Cloud (see lines 11 and 12). Cloud storage services such as OpenStack Swift, Amazon Object Store use the concept of a bucket or a container to store a file. This is why *src_container* and *dest_container* along with *src_filename* and *dest_filename* are given in the *GetCloudFile* function to identify a specific file in the Cloud. The algorithm then compares the hash value of both files and increments *ComparisonCounter*. If all the files in both workflows are the same, *ComparisonCounter* should be equal to *FileCounter*, which counts the number of files produced by a workflow. Thus, it confirms that the workflows are repeated successfully. Otherwise, the algorithms returns false if both these counters are not equal.

---

**Algorithm 3** Compare Workflow Outputs Algorithm

**Require:** $srcWfID$ : Source Workflow ID.
$destWfID$ : Destination Workflow ID

1: **procedure** COMPAREWORKFLOWOUTPUTS($srcWfID$, $destWfID$)
2:   $srcWorkflowJobs \leftarrow$ GETWORKFLOWJOBS($srcWfID$)
3:   $destWorkflowJobs \leftarrow$ GETWORKFLOWJOBS($destWfID$)
4:   $FileCounter \leftarrow 0$
5:   $ComparisonCounter \leftarrow 0$
6:   **for all** $jobfiles \in srcWorkflowJobs$ **do**
7:     $src\_container \leftarrow jobfiles.container\_name$
8:     $src\_filename \leftarrow jobfiles.file\_name$
9:     $dest\_container \leftarrow destWorkflowJobs[jobfiles.jobname]$
10:    $dest\_filename \leftarrow destWorkflowJobs[jobname].file\_name$
11:    $src\_cloud\_file \leftarrow$ GETCLOUDFILE($src\_container$, $src\_filename$)
12:    $dest\_cloud\_file \leftarrow$ GETCLOUDFILE($dest\_container$, $dest\_filename$)
13:    $FileCounter \leftarrow FileCounter + 1$
14:    **if** $src\_cloud\_file.hash = dest\_cloud\_file.hash$ **then**
15:      $ComparisonCounter \leftarrow ComparisonCounter + 1$
16:   **if** $FileCounter = ComparisonCounter$ **then**
17:    **return** True
18:   **return** False

---

Figure 4: Pseudocode to compare outputs produced by two given workflows.

## 5 RESULTS AND DISCUSSION

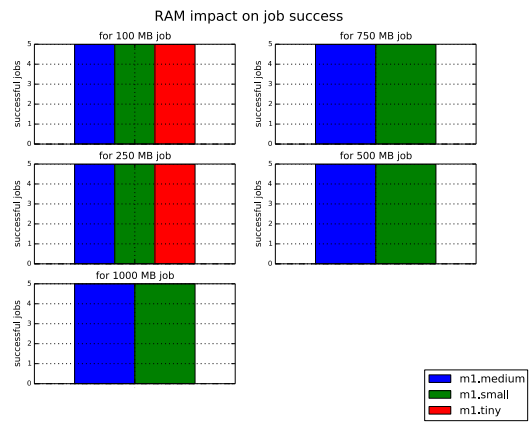

Figure 5: Cloud resource's RAM configuration impact on job success.

To demonstrate the affect of Cloud resource configuration requirement on job failure rate i.e. RAM, a basic memory-consuming job is written in python. This confirms the presented argument in favour of collecting Cloud resource configuration discussed in section 1 and also in section 3. The result in Figure 5 shows that jobs fail if required RAM (hardware) requirement is not fulfilled. Three resource configurations (a) m1.tiny, (b) m1.small and (c) m1.medium, each with 512 MB, 2048 MB and 4096 MB RAM respectively, were used for this experiment. Each job is executed at least 5 times with a given memory requirement on each resource configuration. The following result confirms that as soon as job's memory requirement approaches to 500MB, job starts failing on Cloud resource with m1.tiny configuration.

Table 1: Cloud infrastructure mapped to the jobs of workflow with ID 114.

| wfID | Host IP | nodename | Flavour Id | minRAM (MB) | minHD (GB) | vCPU | Image name | Image id |
|---|---|---|---|---|---|---|---|---|
| 114 | 172.16.1.49 | osdc-vm3.novalocal | 2 | 2048 | 20 | 1 | wf_peg_repeat | f102960c- 557c-4253-8277-2df5ffe3c169 |
| 114 | 172.16.1.98 | mynode.novalocal | 2 | 2048 | 20 | 1 | wf_peg_repeat | 102960c- 557c-4253-8277-2df5ffe3c169 |

Table 2: Cloud infrastructure information of repeated workflow (wfIDs: 117 and 122) after repeating workflow 114.

| wfID | Host IP | nodename | Flavour Id | minRAM (MB) | minHD (GB) | vCPU | Image name | Image id |
|---|---|---|---|---|---|---|---|---|
| 117 | 172.16.1.183 | osdc-vm3-rep.novalocal | 2 | 2048 | 20 | 1 | wf_peg_repeat | f102960c- 557c-4253-8277-2df5ffe3c169 |
| 117 | 172.16.1.187 | mynode-rep.novalocal | 2 | 2048 | 20 | 1 | wf_peg_repeat | f102960c- 557c-4253-8277-2df5ffe3c169 |
| 122 | 172.16.1.114 | osdc-vm3-rep.novalocal | 2 | 2048 | 20 | 1 | wf_peg_repeat | f102960c- 557c-4253-8277-2df5ffe3c169 |
| 122 | 172.16.1.112 | mynode-rep.novalocal | 2 | 2048 | 20 | 1 | wf_peg_repeat | f102960c- 557c-4253-8277-2df5ffe3c169 |

Table 3: Comparing outputs produced by workflows 114 (original workflow) and 117 (repeated workflow).

| Job | WF ID | Container Name | File Name | MD5 Hash |
|---|---|---|---|---|
| Split | 114 | wfoutput123011 | wordlist1 | 0d934584cbc124eed93c4464ab178a5d |
| Split | 117 | wfoutput125819 | wordlist1 | 0d934584cbc124eed93c4464ab178a5d |
| Split | 114 | wfoutput123011 | wordlist2 | 1bc6ffead85bd62b5a7a1be1dc672006 |
| Split | 117 | wfoutput125819 | wordlist2 | 1bc6ffead85bd62b5a7a1be1dc672006 |
| Analysis 1 | 114 | wfoutput123011 | analysis1 | 494f24e426dba5cc1ce9a132d50ccbda |
| Analysis 1 | 117 | wfoutput125819 | analysis1 | 494f24e426dba5cc1ce9a132d50ccbda |
| Analysis 2 | 114 | wfoutput123011 | analysis2 | 127e8dbd6beffdd2e9dfed79d46e1ebc |
| Analysis 2 | 117 | wfoutput125819 | analysis2 | 127e8dbd6beffdd2e9dfed79d46e1ebc |
| Merge | 114 | wfoutput123011 | merge_output | d0bd408843b90e36eb8126b397c6efed |
| Merge | 117 | wfoutput125819 | merge_output | d0bd408843b90e36eb8126b397c6efed |

For workflow based testing, a Python based prototype has been developed using Apache Libcloud (Apache Libcloud – http://libcloud.apache.org) a library to interact with the Cloud middleware. The presented evaluation of the prototype is very basic currently, however, as this work progresses further a full evaluation will be conducted. To evaluate this prototype, a 20 core Cloud infrastructure is acquired from Open Science Data Cloud (OSDC) organisation (https://www.opensciencedatacloud.org/). This Cloud infrastructure uses OpenStack middleware

(openstack.org) to provide the infrastructure-as-a-Service capability. A small Condor cluster of three virtual machines is also configured. In this cluster, one machine is a master node, which is used to submit workflows, and the remaining two are compute nodes. These compute nodes are used to execute workflow jobs. Using the Pegasus APIs, a basic *wordcount* workflow application composed of four jobs is written. This workflow has both control and data dependencies (Ramakrishnan 2010) among its jobs, which is a common characteristic in scientific workflows. The first job (*Split* job) takes a text file and splits it into two files of almost equal length. Later, two jobs (*Analysis* jobs), each of these takes one file as input, and then calculates the number of words in the given file. The fourth job (*merge* job) takes the outputs of earlier analysis jobs and calculates the final result i.e. total number of words in both files.

This workflow is submitted using Pegasus. The wfID assigned to this workflow is 114. The collected Cloud resource information is stored in database. Table I. shows the provenance mapping records in the Provenance Store for this workflow. The collected information includes the *flavour* and *image* (*image name* and *Image id*) configuration parameters. The *Image id* uniquely identifies an OS image hosted on the Cloud and this image contains all the software or libraries used during the job execution. As an image contains all the required libraries of a job, the initial prototype does not extract the installed libraries information from the virtual machine at the moment for workflow repeatability purpose. However, this can be done in future iterations to enable the proposed approach to reconfigure a resource at runtime on the Cloud.

The repeatability of the workflow achieved using the proposed approach (discussed in section 4.2) is also tested. The prototype is requested to repeat the workflow with wfID 114. It first collects the resource configuration from the database and provisions resources on the Cloud infrastructure. It was named *mynova.novalocal* in original workflow execution as shown in Table I. From Table II, one can assess that similar resources have been re-provisioned to repeat the workflow execution because the RAM, Hard disk, CPUs and image configurations are similar to the resources used for workflow with wfID 114 (as shown in Table 1). This preliminary evaluation confirms that the similar resources on the Cloud can be re-provisioned with the Cloud-aware provenance information collected using the proposed approach (discussed previously in section 4). Table 2 shows two repeated instances of original workflow 114.

The other aspect to evaluate the workflow reproducibility (as discussed in section 3) is to verify the outputs produced by both workflows. This has been achieved using the algorithm presented in Figure 4. Four jobs in both the given workflows i.e. 114 and 117 produce the same number of output files (see Table 3). The *Split* job produces two output files i.e. *wordlist1* and *wordlist2*. Two analysis jobs, *Analysis1* and *Analysis2*, consume the wordlist1 and wordlist2 files, and produce the *analysis1* and *analysis2* files respectively. The merge job consumes the *analysis1* and *analysis2* files and produces the *merge_output* file. The hash values of these files are shown in the MD5 Hash column of the Table 3, here both given workflows are compared with each other. For instance, the hash value of *wordlist1* produced by the *Split* job of workflow 117 is compared with the hash value of *wordlist1* produced by the *Split* job of workflow 114. If both the hash values are same, the algorithm will return true. This process is repeated for all the files produced by both workflows. The algorithm confirms the verification of workflow outputs if the corresponding files in both workflows have the same hash values. Otherwise, the verification process fails.

# 6 CONCLUSION AND FUTURE DIRECTION

In this paper, the motivation and the issues related to workflow repeatability due to workflow execution on the Cloud infrastructure have been identified. The dynamic nature of the Cloud makes provenance capturing of workflow(s) and their underlying execution environment(s) and their repeatability a difficult challenge. A workflow reproducibility model (discussed in section 3) has been presented after analysing the literature and workflow execution scenario on Cloud environment. A proposed architecture has been presented that can augment the existing workflow provenance with the information of the Cloud infrastructure. Combining these two can assist in re-provisioning the similar execution environment to reproduce a workflow execution. The Cloud infrastructure information collection mechanism has been presented in this paper in section 4.1. This mechanism iterates over the workflow jobs and establishes mappings with the resource information available on the Cloud. This job to Cloud resource mapping can then be used to repeat a workflow. The process of repeating a workflow execution with the proposed approach has been discussed in section 4.2. In this paper, the workflow repeatability is verified by comparing the outputs produced by the workflows. An algorithm has been discussed in section 4.3 (see Figure 4) that compares the outputs produced by two given workflows. A prototype was developed for evaluation and the results showed that the proposed approach can create a similar execution infrastructure i.e. same resource configuration on the Cloud using the Cloud-aware provenance information (as discussed in section 4) to reproduce a workflow execution. In future, the proposed approach will be extended and a detailed evaluation of the proposed approach will be conducted. Different performance matrices such as the impact of the proposed approach on workflow execution time, impact of different resource configuration on workflow performance, and total resource provision time will also be measured. In future, more emphasis will be given to the mechanisms to incorporate the workflow provenance comparison (as discussed in section 3) to verify workflow repeatability.

## ACKNOWLEDGEMENTS

This research work has been funded by a European Union FP-7 project, N4U – neuGrid4Users. This project aims to assist the neuro-scientific community in analysing brain scans using workflows and distributed infrastructure (Grid and Cloud) to identify biomarkers that can help in diagnosing the Alzheimer disease. Besides this, the support provided by OSDC by offering a free Cloud infrastructure of 20 cores is highly appreciated. Such public offerings can really benefit research and researchers who are short of such resources.